# Complexity in atoms:
# An approach with a new analytical density

Jaime Sañudo* and Ricardo López-Ruiz**

*Departamento de Física, Facultad de Ciencias,*

*Universidad de Extremadura, E-06071 Badajoz, Spain*

**DIIS and BIFI, Facultad de Ciencias,*

*Universidad de Zaragoza, E-50009 Zaragoza, Spain*

# **Abstract**

In this work, the calculation of complexity on atomic systems is considered. In order to unveil the increasing of this statistical magnitude with the atomic number due to the relativistic effects, recently reported in [A. Borgoo, F. De Proft, P. Geerlings, K.D. Sen, Chem. Phys. Lett., **444** (2007) 186], a new analytical density to describe neutral atoms is proposed. This density is inspired in the Tietz potential model. The parameters of this density are determined from the normalization condition and from a variational calculation of the energy, which is a functional of the density. The density is non-singular at the origin and its specific form is selected so as to fit the results coming from non-relativistic Hartree-Fock calculations. The main ingredients of the energy functional are the non-relativistic kinetic energy, the nuclear-electron attraction energy and the classical term of the electron repulsion. The relativistic correction to the kinetic energy and the Weizsacker term are also taken into account. The Dirac and the correlation terms are shown to be less important than the other terms and they have been discarded in this study. When the statistical measure of complexity is calculated in position space with the analytical density derived from this model, the increasing trend of this magnitude as the atomic number increases is also found.

\*   Electronic address: jsr@unex.es

\*\*  Electronic address: rilopez@unizar.es

# 1. Introduction

Despite a universal measure of complexity is yet unknown, the application of complexity to atomic systems is a topic of great scientific interest [1]. In a recent work [2], it has been shown the marked influence that the consideration of relativistic effects in atoms has on a statistical measure of complexity, $C_{LMC}$, defined in Refs. [3, 4]. The main ingredient for the calculation of the complexity is the electron density in the atom, $n(\vec{r})$. Different electron densities have been proposed through the years [5, 6]. Using the non-relativistic Hartree-Fock wave functions, the analysis of the electronic structural complexity for atoms with atomic number $Z = 2-54$ has shown an slight increase of complexity when the atomic number increases [1, 7, 8]. When the Dirac-Fock relativistic wave functions are used, the increasing trend of the complexity with the atomic number, $Z = 1-103$, is enhanced [2]. Knowing these facts, we can wonder if the inclusion of the relativistic terms in simpler atomic models also causes this kind of behaviour.

In this work, our aim is to unveil this possibility in a modified Thomas-Fermi model [6, 9]. In order to reach this goal, this study is twofold: first, to obtain an analytical density that incorporates the relativistic effects in a perturbative manner and that is well behaved at the origin. This is presented in Section 2. And second, to use this density to check the influence of the relativistic effects on the complexity. This calculation is performed in Section 3. The conclusions are included in Section 4.

## 2. The Analytical Density

Hohenberg and Kohn [10] show that the ground-state energy of a quantum-mechanical system can be written as a density functional, $E[n]$. However, it is not established the specific form of that functional. Thus, for our purposes, the ingredients of the energy functional for the atom are presented in Section 2.1. In Section 2.2, an analytical expression for the density of the atom is justified. All the parameters of this density are determined by using the minimization of the energy and the normalization condition, and by fitting the density at the origin to Hartree-Fock calculations.

### 2.1 The Functional E[*n*]

The total energy, $E$, for a point-like nucleus of atomic number $Z$ surrounded by an electron cloud, can be obtained from the energy-density functional [11, 12],

$$E[n] = E_{kin}[n] + E_{eN}[n] + E_{ee}[n] \quad , \tag{1}$$

where $E_{kin}[n]$ is the kinetic energy (whose expression is given below), $E_{eN}[n]$ is the electron-nuclear attraction energy

$$E_{eN}[n] = -Ze^2 \int \frac{n(\vec{r})}{r} d\vec{r} \quad , \tag{2}$$

$e$ being the charge of the electron, and $E_{ee}[n]$ is the electron-electron repulsion energy. This last term can be divided into two parts,

$$E_{ee}[n] = J[n] + K[n] \quad , \tag{3}$$

where $J[n]$ is the classical Coulomb repulsion between the electrons

$$J[n] = \frac{e^2}{2} \iint \frac{n(\vec{r})n(\vec{r}')}{|\vec{r}-\vec{r}'|} d\vec{r}d\vec{r}' \quad , \tag{4}$$

and $K[n]$ is called the exchange-correlation energy, that, in turn, can also be separated into

$$K[n] = K_D^{NR}[n]\{1+\cdots\} + K_{Corr}^{NR}[n]\{1+\cdots\} \quad . \tag{5}$$

The first term in (5) is the exchange energy, and here, we adopt for it the non-relativistic (*NR*) homogeneous electron gas approximation of Dirac [11, 12],

$$K_D^{NR}[n] = \lambda_D e^2 \int n(\vec{r})^{4/3} d\vec{r} \quad , \tag{6}$$

where $\lambda_D = -\frac{3}{4}\left(\frac{3}{\pi}\right)^{1/3}$. The second term in (5) is the *NR* correlation energy obtained by Ceperley and Adler [13], and precisely fitted by Barbiellini-Amidi [14] by means of the expression

$$K_{Corr}^{NR}[n] = \lambda_{Cep} \frac{e^2}{a_B^{1/2}} \int n(\vec{r})^{7/6} d\vec{r} \quad , \tag{7}$$

with $a_B = \frac{\hbar^2}{me^2}$ the Bohr radius, $m$ the electron mass, and $\lambda_{Cep} = -0.0635\left(\frac{4\pi}{3}\right)^{1/6}$. It should be noted that in Eq. (5) the ellipsis inside the curly brackets is meaning the relativistic corrections [12] that we are not considering here.

For the kinetic energy we adopt the first two terms of the gradient expansion [11, 12],

$$E_{kin}[n] = T_0[n] + T_2^{NR}[n]\{1+\cdots\} + \cdots \quad , \tag{8}$$

where the quantity

$$T_2^{NR}[n] = \frac{1}{9}\lambda_W \frac{\hbar^2}{m} \int \frac{(\vec{\nabla}n)^2}{n} d\vec{r} \qquad (9)$$

is one-ninth the inhomogeneity correction introduced by Weizsacker, with $\lambda_W = \frac{1}{8}$, and the ellipsis inside the curly brackets in Eq. (8) has the same meaning as said above. On the other hand, the quantity $T_0[n]$ can be split in

$$T_0[n] = T_0^{NR}[n] + T_0^R[n] \, , \qquad (10)$$

where the term $T_0^{NR}[n]$ is the Thomas-Fermi kinetic energy [6,9] given by

$$T_0^{NR}[n] = \frac{3}{10}(3\pi^2)^{2/3} \frac{\hbar^2}{m} \int n^{5/3} d\vec{r} \, , \qquad (11)$$

and $T_0^R[n]$ is the relativistic corrections, to the first order, for the homogeneous electron gas [6],

$$T_0^R[n] = -\frac{15}{280}(3\pi^2)^{4/3} \frac{\hbar^2}{m} \alpha^2 a_B^2 \int n^{7/3} d\vec{r} + o(\alpha^4) \, , \qquad (12)$$

$\alpha = \frac{e^2}{\hbar c}$ being the fine structure constant.

The density we are looking for is the solution that minimizes the energy $E[n]$, given in Eq. (1), taking into account the constrain for the normalization of the density,

$$N = \int n(\vec{r}) d\vec{r} \, , \qquad (13)$$

where $N$ denotes the number of electrons. Here, we consider the neutral atom, $N = Z$. Within the limits of the problem, we proceed to do that in the next section.

## 2.2 The Density and the Minimization of the Energy

Let us start by recalling the well known Thomas-Fermi solution for Eq. (1). This is obtained when the energy is minimized by avoiding the relativistic correction $(\alpha \equiv 0)$, the Weizsacker term $(\lambda_W \equiv 0)$ and the exchange- correlation energy $(\lambda_D \equiv 0, \lambda_{Cep} \equiv 0)$ with the constrain given by (13). This solution, namely the Thomas-Fermi density for neutral atoms with a point-like nucleus at zero temperature [6, 9], is $n(x) = \frac{32}{9\pi^3} \frac{Z^2}{a_B^3} \left(\chi(x)/x\right)^{3/2}$, with $b = \left(\frac{9\pi^2}{128}\right)^{1/3} Z^{-1/3} a_B$ and $r = bx$. The potential $\chi(x)$ can not be found analytically. The Thomas-Fermi model has also been studied in the relativistic case for non point-like nucleus in [15] and at finite temperature in [16], and also in these cases the solution can be only obtained numerically. Tietz [17] proposed an analytic expression for that potential, $\chi(x) = \frac{1}{(1+\beta x)^2}$, that fits very well the exact Thomas-Fermi solution in the range $0 \leq x \cong 10$ [18]. The parameter $\beta$ can be determined by minimizing the energy or by using Eq. (13), but, evidently, it is not possible to reach the fulfilment of both conditions with only one parameter.

Inspired in the Tietz potential model, we propose the following density:

$$n(x) = \frac{32}{9\pi^3} \frac{Z^2}{a_B^3} \frac{\varepsilon}{(1+\beta x)^3 (x+\gamma)^{3/2}}, \qquad (14)$$

to describe neutral atoms, with three parameters, $\gamma$, $\beta$ and $\varepsilon$ to be determined. The parameter $\gamma$ allows the density to be non singular at the origin. Kato [19] proved the important result that in a quantum system the density should fulfil the condition

$\frac{dn(r=0)}{dr} = -2\frac{Z}{a_B}n(r=0)$. With the density given in Eq. (14), the last condition

implies that $\gamma = \frac{3}{2}\left(\frac{4}{3\pi}\right)^{2/3} Z^{-2/3} + o(Z^{-4/3})$. Therefore, we choose for $\gamma$ the form

$$\gamma = \frac{3}{2}\left(\frac{4}{3\pi}\right)^{2/3} \lambda Z^{-2/3}, \qquad (15)$$

where $\lambda$ is a new parameter not very far from 1 that will be determined latter.

The parameter $\varepsilon$ in the density (14) is fixed by means of the normalization condition (13). We obtain

$$\varepsilon = \frac{\beta^{3/2}}{F(\gamma \cdot \beta)}, \qquad (16)$$

where

$$F(\gamma \cdot \beta) = \frac{1}{4}\left\{ \frac{1 - 8\gamma \cdot \beta - 8(\gamma \cdot \beta)^2}{(1-\gamma \cdot \beta)^{7/2}}\frac{\pi}{2} + \frac{(\gamma \cdot \beta)^{1/2}(14\gamma \cdot \beta + 1)}{(1-\gamma \cdot \beta)^3} + \frac{(-1 + 8\gamma \cdot \beta + 8(\gamma \cdot \beta)^2)}{(1-\gamma \cdot \beta)^{7/2}}arctan\sqrt{\frac{\gamma \cdot \beta}{1-\gamma \cdot \beta}} \right\}. \qquad (17)$$

For $\gamma \cdot \beta << 1$, the behaviour of $F(\gamma \cdot \beta)$ is

$$F(\gamma \cdot \beta << 1) = \frac{\pi}{8} - \frac{9\pi}{16}\gamma \cdot \beta + o((\gamma \cdot \beta)^2). \qquad (18)$$

Therefore, the only parameter to be determined at this point is $\beta$. This will be found with the minimization of the energy.

We substitute Eqs. (16) and (18) in the density (14), and plug it in Eq. (1),

$$E[n] = T_0^{NR}[n] + T_2^{NR}[n] + T_0^R[n] + E_{eN}[n] + J[n] + K_D^{NR}[n] + K_{Corr}^{NR}[n]. \qquad (19)$$

For $\gamma \cdot \beta << 1$, the power expansion of the different terms is:

$$\frac{T_0^{NR}}{e^2/a_B} = 2.764 Z^{7/3}\beta^2 \left\{1 - 6.209(\gamma\cdot\beta)^{1/2} + o((\gamma\cdot\beta))\right\}, \qquad (20)$$

$$\frac{T_2^{NR}}{e^2/a_B} = 0.8663\lambda_W Z^{5/3}\gamma^{-1/2}\beta^{3/2}\left\{1 + o((\gamma\cdot\beta)^{1/2})\right\}, \qquad (21)$$

$$\frac{T_0^R}{e^2/a_B} = -2.582(\alpha Z)^2 Z^{5/3}\gamma^{-1/2}\beta^{3/2}\left\{1 + o((\gamma\cdot\beta)^{1/2})\right\}, \qquad (22)$$

$$\frac{E_{eN}}{e^2/a_B} = -3.389 Z^{7/3}\beta\left\{1 - 3.395(\gamma\cdot\beta)^{1/2} + o((\gamma\cdot\beta))\right\}, \qquad (23)$$

$$\frac{J}{e^2/a_B} = 0.4735 Z^{7/3}\beta\left\{1 + o((\gamma\cdot\beta))\right\}, \qquad (24)$$

$$\frac{K_D^{NR}}{e^2/a_B} = -0.5631\lambda_D Z^{5/3}\beta\left\{1 + o((\gamma\cdot\beta)^{1/2})\right\}, \qquad (25)$$

$$\frac{K_{Corr}^{NR}}{e^2/a_B} = 0.6410\lambda_{Cep} Z^{4/3}\beta^{1/2}\left\{1 + o((\gamma\cdot\beta)^{1/2})\right\}. \qquad (26)$$

We advance that it will be obtained that $\beta \approx 0.5$, then the former expressions (20-26) are power expansions in $\gamma$. Keeping in mind that $\gamma$ is proportional to $Z^{-2/3}$, see Eq. (15), then Eqs. (20-26) display the typical development of the energy terms in decreasing powers of $Z^{1/3}$ [6, 9]. So, for $T_0^{NR}$, $E_{eN}$ and $J$ the dominant power is $Z^{7/3}$, for both $T_2^{NR}$ and $T_0^R$ the dominant power is $Z^2$. In this last term, $T_0^R$, the factor $\alpha Z$ has been taken as a constant of order 1 due to the fact that the relativistic effects in which we are interested are mainly important when Z>>1. The term $K_D^{NR}$ goes as $Z^{5/3}$, whereas $K_{Corr}^{NR}$ as $Z^{4/3}$.

Now, the parameter β can be obtained by minimizing the energy, $\frac{dE}{d\beta} = 0$. In order to do zero all the coefficients of the powers of Z in that derivative, we find that β must be expressed in decreasing powers of $Z^{1/3}$ as

$$\beta = \beta_0 \{1 + a\gamma^{1/2} + o(\gamma)\}. \tag{27}$$

Equalling to zero the coefficient of the power $Z^{7/3}$, $\beta_0$ is obtained

$$\beta_0 = 0.5272. \tag{28}$$

In the same manner, from equalling to zero the coefficient of the term in $Z^2$, we find that

$$a = 1.336 + \frac{1}{\lambda}\{1.139(\alpha Z)^2 - 0.3837\lambda_W\}. \tag{29}$$

The last parameter to be fixed in our density model is λ. From (14), the behaviour of our density model at origin reads

$$\frac{n(0)a_B^3}{Z^3} = \frac{32}{9\pi^3} \frac{\varepsilon}{(Z\gamma^{3/2})}. \tag{30}$$

Substituting ε and γ given by Eqs. (15) and (16) into (30), and taking into account (27)-(29), we find that the right-hand side of (30) is only dependent of the parameter λ. After taking $\alpha \equiv 0$, we can fit the right-hand side of Eq. (30) with the non-relativistic Hartree-Fock calculation of $n(0)a_B^3/Z^3$ due to Fischer [20, 21]. The value obtained for λ is

$$\lambda = 0.3960 - 0.3388Z^{-1/3} + 1.714Z^{-2/3}. \tag{31}$$

This fit is shown in Fig.1 where we have represented $\frac{n(0)a_B^3}{Z^3}$ versus *Z*. The continuous line shows Eq. (30) with λ given by Eq. (31), and the dots represent the non-relativistic Hartree-Fock calculation due to Fischer. Let us note that the fit is excellent.

We should indicate at this point that the procedure becomes cumbersome if one tries to perform the calculations for lower orders than $Z^2$. Hence, we have cut the development of β to the zero and first order in powers of $\gamma^{1/2}$ in Eq. (27). Also, let us note that the exchange-correlations terms given in Eqs. (25-26) are of order lower than $Z^2$, and therefore, negligible until the order of our approximation. A further calculation to obtain more terms could be reported elsewhere.

Let us remark the goodness of this density (14) to calculate the energy of a non-relativistic neutral atom. According to Eq. (19), this yields $E = 0.7684 Z^{7/3} + o(Z^2)$, where it has been taken the order zero in the development of the parameters: ε = 0.9748, β = 0.5272 and $\gamma = 0.3355 Z^{-2/3}$. Knowing that the exact Thomas-Fermi energy [6, 9] is $E^{TF} = 0.7688 Z^{7/3}$ and the energy derived from the original Tietz density gives $E^{Tietz} = 0.7682 Z^{7/3}$ (obtained by doing ε = 1 and γ = 0, whereas β is 0.5632 when the normalization condition is used), the precision of our calculation is notable. Furthermore, the density (14) here proposed allows us to incorporate the relativistic effects with the possibility to see their influence in different magnitudes, such as we will show with the complexity in the next section.

# 3. Statistical Complexity

Now, we calculate with the density (14) several magnitudes related with statistical complexity. As it has been reported in [2] when Dirac-Fock relativistic wave functions are used, here we also show the influence of the relativistic effects in those magnitudes, in particular the increasing trend of the complexity with the atomic number.

The measure of complexity $C$, the so-called *LMC* complexity [3,4], is defined as

$$C_{LMC} = H \cdot D \quad , \tag{32}$$

where $H$ represents the information [4, 22] content of the system

$$H = \frac{1}{2\pi e} e^{2S_r/3} \quad , \tag{33}$$

$S_r$ being the Shannon information entropy [23] in position space,

$$S_r = -\int \hat{n}(\vec{r}) log\left(a_B^3 \hat{n}(\vec{r})\right) d\vec{r} \quad , \tag{34}$$

and $D$ is calculated as the density expectation value [3, 4]

$$D = a_B^3 \int \hat{n}^2(\vec{r}) d\vec{r} \quad . \tag{35}$$

In Eqs. (34-35) the electron density is normalized to unity, therefore $\hat{n} \equiv \frac{n}{Z}$, and $n$ is given in Eq. (14).

Let us recall at this point that $C$ has been quantified in different contexts (see Ref. [24] and references therein). It has been shown that $C$ is a useful indicator to successfully discern many situations regarded as complex in systems out of equilibrium [24]. Thus, $C$ identifies the entropy or information $H$ stored in a system and its disequilibrium $D$, that in the discrete case is the distance from its actual state to another probability distribution of equilibrium, as the two basic ingredients for calculating the complexity of a system. In the case of a continuous support in the distributions, the magnitudes are redefined as indicated in Eqs. (33-35) (see [4] for a detailed discussion).

In Fig. 2, $D$ versus $Z$ is plotted. The continuous line represents the calculation including the relativistic correction to the kinetic energy, it is to say, the influence of the term given in Eq. (12). The dotted line is a similar calculation but omitting the relativistic correction, that is, when $\alpha \equiv 0$. We can see the importance of taking into account the relativistic correction to the kinetic energy. This result qualitatively agrees with that of the Ref. [2], where Dirac-Fock relativistic wave functions are used.

In Fig. 3, we plot $H$ versus $Z$. The continuous line and the dotted line represent, as before, the relativistic and non-relativistic calculations, respectively. Qualitatively speaking, the relativistic influence here is slightly bigger than in the results presented in Ref. [2].

Finally, in Fig. 4, we show $C$ versus $Z$. The meaning of continuous line and the dotted line are the same as before. Let us observe that the influence of the relativistic effects here are qualitatively less important than those obtained in Ref. [2].

All these results can be also compared with the non-relativistic calculations performed in Ref. [25]. There, several statistical complexities were obtained by using the universal electron density found by Gáspár [26], and, in general, an increasing trend of these magnitudes with increasing atomic number was reported. This tendency is also found in Fig. 4 with the difference that our approach incorporates the influence of relativistic corrections to the kinetic energy.

## 4. Conclusions

In this work, a new analytical density to describe neutral atoms has been proposed. The specific form of this density is inspired in the Tietz potential model. This density, which is not singular at the origin, has three parameters that are fixed by means of three constraints: the normalization condition, the minimization of the energy as a functional of the density and the fit of the density at the origin with non-relativistic Hartree-Fock calculations. We have used as ingredients for the energy functional the non-relativistic kinetic energy, the attractive nuclear-electron energy, the classical repulsive electron interaction, the relativistic correction to the kinetic energy, the Weizsacker term, and the Dirac and correlation terms. Up to order $Z^2$ in the energy functional, the last two terms, namely Dirac and correlation ones, are negligible and therefore they have been omitted. After minimizing the energy functional, we have obtained the density.

The calculation of the energy with this density yields $E = 0.7684Z^{7/3} + o(Z^2)$. This result is comparable with the exact Thomas-Fermi energy, $E^{TF} = 0.7688Z^{7/3}$, and with the energy derived from the original Tietz density, $E^{Tietz} = 0.7682Z^{7/3}$. Moreover, this density has allowed us to see the qualitative influence of the relativistic effects in the statistical complexity.

Thus, we have calculated with this new density different statistical magnitudes: $H$, which denotes the Shannon information, $D$, which represents the disequilibrium, and $C_{LMC}$ that is a statistical measure of complexity. We have made manifest the qualitative influence of the relativistic corrections in all these magnitudes in agreement to the behaviour found in Ref. [2], where the increasing trend of the complexity with the atomic number was put in evidence. Thus, Borgoo *et al*. [2] used the Dirac-Fock relativistic wave functions to unmask this behaviour, whereas our result has been obtained in an analytical manner.

Let us conclude by saying that some simple analytical approaches can reproduce the qualitative behaviour of different statistical magnitudes in atomic systems, and so, to help our physical intuition in trying to discern complexity at a quantum level.

# Acknowledgements

The authors acknowledge some financial support by grant DGICYT-FIS2005-06237.

# Figure Captions

Fig. 1.- Atomic density at the origin as a function of $Z$. The continuous line represents our density with $\lambda$ given by Eq. (31) after taking $\alpha = 0$. The dots are the non-relativistic Hartree-Fock calculations of Fischer (see the text).

Fig. 2.- The disequilibrium, $D$, versus the atomic number, $Z$, as given in Eq. (35). The dotted line represents the non-relativistic calculation ($\alpha \equiv 0$), whereas the continuous line is the relativistic case (see the text).

Fig. 3.- The Shannon entropy, $H$, versus the atomic number, $Z$, as given in Eq. (33). The comments done in Fig. 2 are also valid here.

Fig. 4.- The LMC complexity measure, $C_{LMC}$, as a function of atomic number, $Z$, as given in Eq. (32). The comments in Fig. 2 are also valid here.

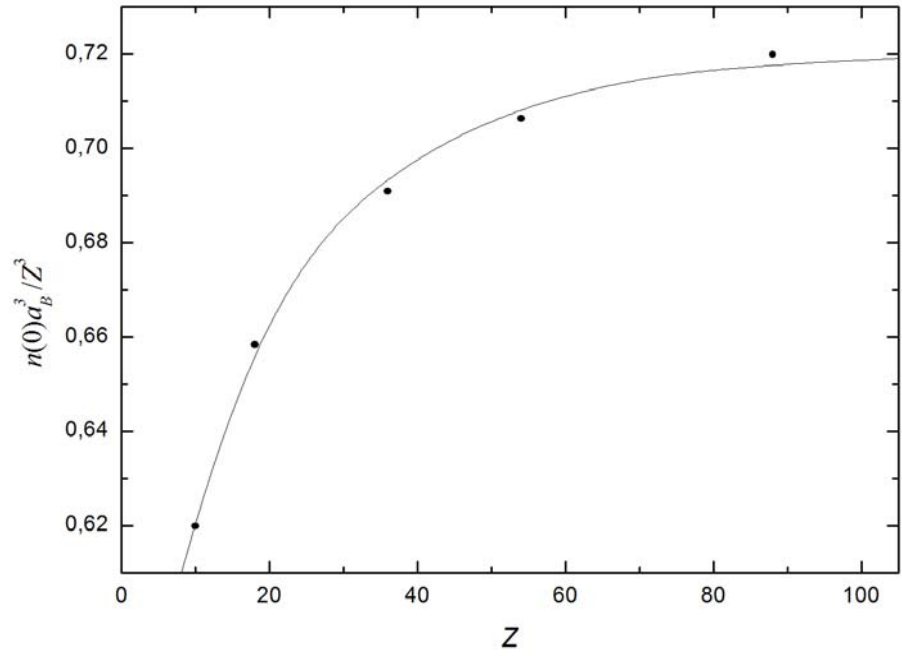

Fig.1

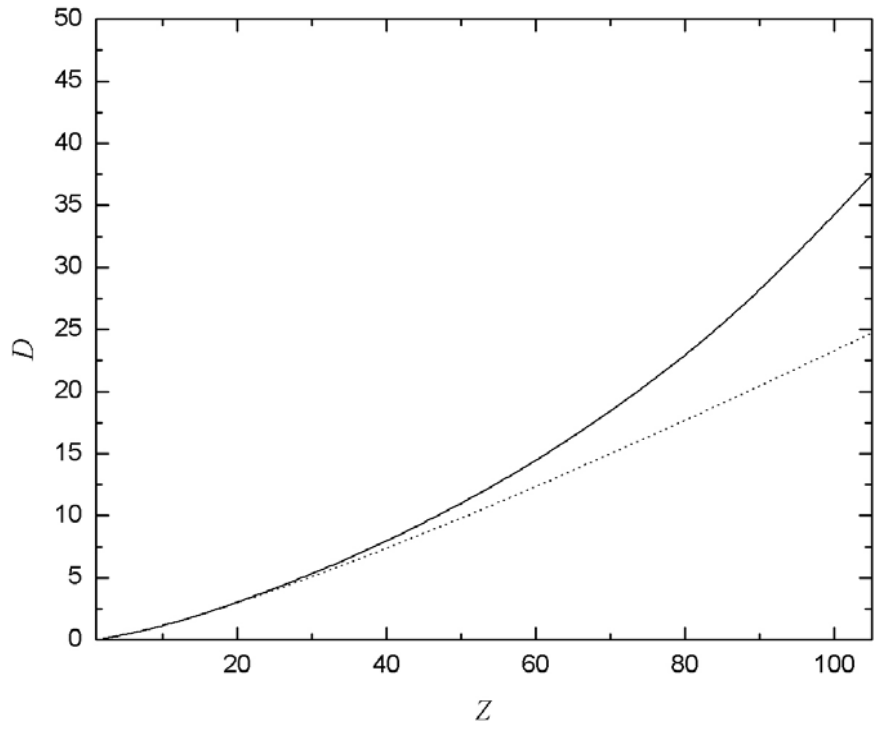

Fig.2

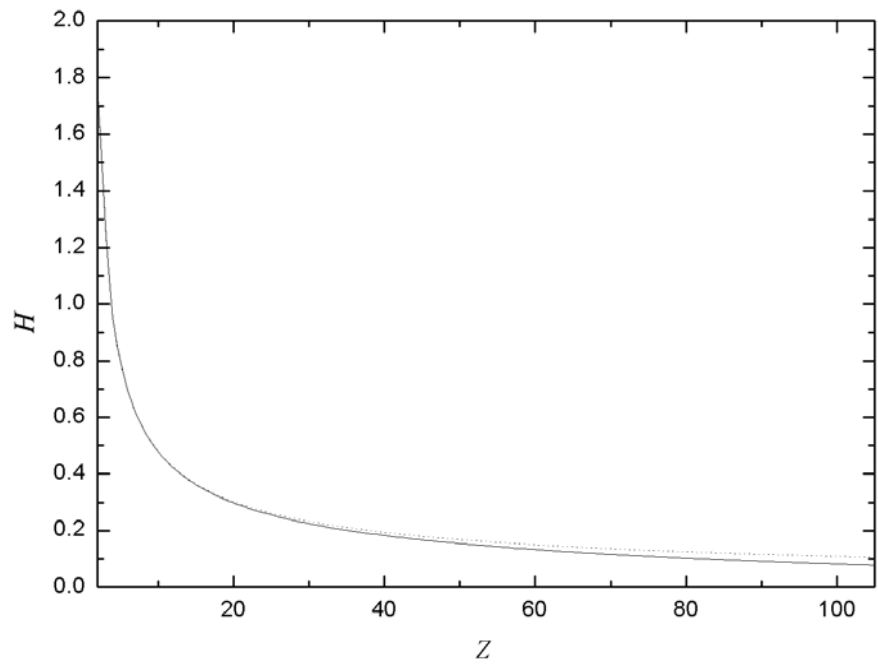

Fig.3

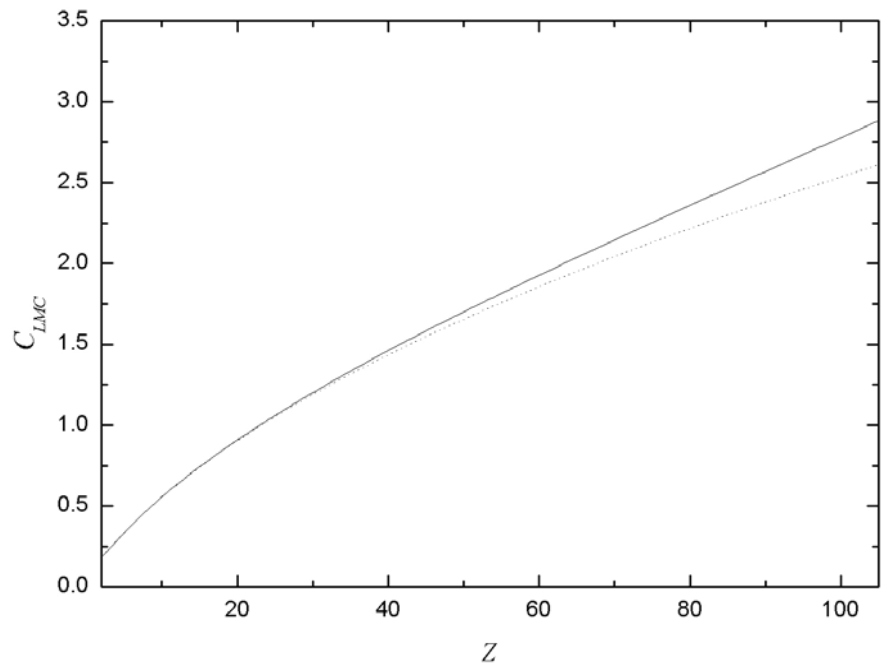

Fig.4